\documentclass[12pt]{article}
\usepackage{epsf}
\usepackage{amsmath}
\usepackage{amssymb}
\usepackage{amsfonts}
\usepackage{mathrsfs}
\usepackage{cite}
\usepackage{bbm}
\usepackage{graphicx}
\usepackage[footnotesize]{caption}
\usepackage{mathrsfs}

\newcommand{\bmat}{\left(\begin{array}}
\newcommand{\emat}{\end{array}\right)}

\def\yzero{\smash{\hbox{$y\kern-4pt\raise1pt\hbox{${}^\circ$}$}}}

\def\beq{\begin{equation}}
\def\eeq{\end{equation}}
\def\beqa{\begin{eqnarray}}
\def\eeqa{\end{eqnarray}}

\def\-{\hphantom{-}}

\def\s2{\frac{1}{\sqrt2}}

\def\beq{\begin{equation}}
\def\eeq{\end{equation}}
\def\beqa{\begin{eqnarray}}
\def\eeqa{\end{eqnarray}}

\def\diag{{\rm diag \,}}
\def\IF{\relax{\rm I\kern-.18em F}}
\def\II{\relax{\rm I\kern-.18em I}}
\def\IP{\relax{\rm I\kern-.18em P}}
\def\IC{\relax\hbox{\kern.25em$\inbar\kern-.3em{\rm C}$}}
\def\IR{\relax{\rm I\kern-.18em R}}

\def\Dsl{\,\raise.15ex\hbox{/}\mkern-13.5mu D} %this one can be subscripted
\def\IZ{Z\kern-.4em  Z}

\def\<{\left\langle}
\def\>{\right\rangle}

\catcode`\@=11
\newdimen\@rotdimen
\newbox\@rotbox

\def\@vspec#1{\special{ps:#1}}%  passes #1 verbatim to the output
\def\@rotstart#1{\@vspec{gsave currentpoint currentpoint translate
   #1 neg exch neg exch translate}}% #1 can be any origin-fixing transformation
\def\@rotfinish{\@vspec{currentpoint grestore moveto}}% gets back in synch
\def\@rotr#1{\@rotdimen=\ht#1\advance\@rotdimen by\dp#1%
   \hbox to\@rotdimen{\hskip\ht#1\vbox to\wd#1{\@rotstart{90 rotate}%
   \box#1\vss}\hss}\@rotfinish}
\def\@rotl#1{\@rotdimen=\ht#1\advance\@rotdimen by\dp#1%
   \hbox to\@rotdimen{\vbox to\wd#1{\vskip\wd#1\@rotstart{270 rotate}%
   \box#1\vss}\hss}\@rotfinish}%
\def\@rotu#1{\@rotdimen=\ht#1\advance\@rotdimen by\dp#1%
   \hbox to\wd#1{\hskip\wd#1\vbox to\@rotdimen{\vskip\@rotdimen
   \@rotstart{-1 dup scale}\box#1\vss}\hss}\@rotfinish}%
\def\@rotf#1{\hbox to\wd#1{\hskip\wd#1\@rotstart{-1 1 scale}%
   \box#1\hss}\@rotfinish}%
\def\rotate{\@ifnextchar[{\@rotate}{\@rotate[l]}}
\def\@rotate[#1]#2{\setbox\@rotbox=\hbox{#2}\@nameuse{@rot#1}\@rotbox}

\catcode`\@=12
\topmargin
-1.5cm
\textwidth
15.5cm
\textheight
23.5cm
\oddsidemargin
0.7cm
\evensidemargin
1.2cm

\begin{document}

%----------------------------------------------------------------------%
%  numbering equations with section number
%----------------------------------------------------------------------%
\makeatletter
\@addtoreset{equation}{section}
\makeatother
\renewcommand{\theequation}{\thesection.\arabic{equation}}
%----------------------------------------------------------------------%
%  title page
%----------------------------------------------------------------------%
\pagestyle{empty}
%\vspace*{1.0in}
\rightline{ IFT-UAM/CSIC-07-35}
%\rightline{\tt hep-ph/yymmnnn}
\vspace{2.5cm}
\begin{center}
\LARGE{ Neutrino  Masses and Mixings \\ from  String Theory Instantons
\\[10mm]}
\large{ S. Antusch$^{1,2}$, L.E. Ib\'a\~nez$^{1}$ and T. Macr\`i$^{1,3}$  \\[6mm]}
\small{
${}^{1}$~Departamento de F\'{\i}sica Te\'orica C-XI
and Instituto de F\'{\i}sica Te\'orica  C-XVI,\\[-0.3em]
Universidad Aut\'onoma de Madrid,
Cantoblanco, 28049 Madrid, Spain \\[2mm]
${}^{2}$~Max-Planck-Institut f\"ur Physik (Werner-Heisenberg-Institut) \\[-0.3em]
 F\"ohringer Ring 6, D-80805 M\"unchen, Germany
\\[2mm]
$^{3}$~Dipartimento di Fisica 'G. Galilei', Universit\`a di Padova\\[-0.3em]
Via Marzolo 8, I-35131 Padua, Italy\\ [7mm]}

\vspace{7mm}

%\small{\bf Abstract} \\[-5mm]
\end{center}
\begin{center}
\small{\bf Abstract} \\[7mm]
\begin{minipage}[h]{16.0cm}
We study possible patterns of neutrino masses and mixings
in string models in which Majorana neutrino masses
are generated by a certain class of string theory instantons
recently considered in the literature. These instantons may generate
either directly the dim=5 Weinberg operator or right-handed
neutrino Majorana masses, both with a certain flavour-factorised form.
A hierarchy of neutrino masses naturally appears from  the
exponentially suppressed contributions  of different instantons.
The flavour structure is controlled by string amplitudes
involving neutrino fields and charged instanton zero modes.
For some simple choices for these amplitudes one finds
neutrino mixing patterns consistent with experimental results.
In particular, we find that a tri-bimaximal mixing pattern is obtained
for simple symmetric values of the string correlators.

\end{minipage}
\end{center}
\newpage
%----------------------------------------------------------------------%
%  Resetting of counters
%----------------------------------------------------------------------%
\setcounter{page}{1}
\pagestyle{plain}
\renewcommand{\thefootnote}{\arabic{footnote}}
\setcounter{footnote}{0}
%----------------------------------------------------------------------%
%  Paper begins
%----------------------------------------------------------------------%

%\end{document}

\section{Introduction}

Recently a new mechanism for the generation of neutrino Majorana masses
in the context of string theory has been pointed out \cite{iu,bcw,crw,isu}.
 Certain string instanton effects can generate
right-handed  neutrino masses from operators of the form
\beq
e^{- { U_M}}\,  {\nu_R\nu_R } {M_{string}} \;.
\label{operon}
\eeq
Here $M_s$ is  the string scale and  $U_M$ is a complex scalar modulus field whose
axion-like imaginary part $Im\,U_M$
gets shifted under a gauged $U(1)_{B-L}$ symmetry in such a way
that the operator (\ref{operon}) is  $U(1)_{B-L}$ gauge invariant.
The size of these masses is of order $\exp(- Re\, U_M)M_s$. Unlike ordinary,
e.g., electroweak  instanton
effects which are of order $\exp(-1/\alpha_2 )$,
these instantons need not be very much suppressed, $Re\,U_M$ is not the inverse
of any SM gauge coupling and may be relatively small. Thus right-handed neutrino masses
may be large, as required phenomenologically.
Furthermore it was noted \cite{isu} that analogous instantons can also generate
a dimension 5 Weinberg operator of the form
\beq
e^{- { U_W}}\, \frac {1}{ {M_s}}  {{\overline H}L}  {{\overline H}L}\;.
\label{operin}
\eeq
This term gives rise directly to left-handed neutrino masses once the Higgs
scalars get a vev.
Both these instanton effects only appear in a restricted class of string
compactifications in which the SM gauge group is extended by a $U(1)_{B-L}$
gauge boson which is massive through a Stuckelberg mass term.
String compactifications in which such instanton mechanism is operative have
been recently discussed in
\cite{iu,bcw,crw,isu}.

In the present paper we make a first phenomenological exploration of the
structure of neutrino masses and mixings obtained from this string
instanton mechanism. In this  analysis we will
 concentrate on a particular class of instantons, those with
internal $Sp(2)$ Chan-Paton (CP)  symmetry which leads to the simplest structure
and appear most often in available instanton searches \cite{isu}.
For such instantons the flavour dependence of both   $\nu_R$-masses and
the Weinberg operator factorises as product of flavour vectors (called
$d_a$ and $c_a$ (a=1,2,3) in the main text
for the $\nu_R$-masses and Weinberg operator respectively). These flavour vectors
$d_a,c_a$ may be in principle computable in terms of the specific underlying
string compactification. This simple flavour structure and the fact that one
expects several different instantons contributing to the amplitude
make it quite natural to obtain a hierarchy of neutrino masses \cite{isu}.

The structure of this paper is as follows. In the next section, section 2, we
present a brief overview of the string instanton mechanism which
is relevant for the generation of neutrino Majorana masses. We discuss how the
operators in Eqs.~(\ref{operon}) and (\ref{operin}) may be generated as well as their flavour structure and the expected size of the neutrino masses.
Turning to the phenomenological analysis in section 3, in section 3.1 we consider the
case in which the Weinberg operator is dominant compared to the see-saw contribution.
We also assume in a first approximation that the large mixing in the
leptonic sector originates in the neutrino mass matrix (and not in the charged
leptons).
In this case the physical neutrino mass matrix is directly obtained from the
discussed instanton effects and the analysis is much easier. We
show that,
if there is a hierarchy of neutrino masses (naturally induced by the 
above-named instanton effects),  then one can obtain a neutrino mixing matrix
consistent with experimental results for certain (not very stringent)
constraints on the values of the flavour vectors $c_a$.
We also show that for flavour vectors $c_a$ along particular
directions one can reproduce, e.g., tri-bimaximal mixing
both for the normal and inverse hierarchy cases.

We then consider the case in which the see-saw contribution to neutrino masses
is dominant in section 3.2. In this case the final result for the physical
neutrino masses depends on the structure of the Dirac mass for the neutrinos.
This makes the
analysis more model-dependent. We consider a simple case in which the Dirac mass
matrix is diagonal. In this case one can obtain e.g. tri-bimaximal mixing
if the flavour vector coefficients  $d_a$ of different contributing instantons
align along certain directions in flavour space.
The case in which both the Weinberg operator and see-saw mechanism are
relevant is briefly discussed in section 3.3. Some final conclusions
and some comments are left to section 4.

\section{Neutrino masses and string instanton effects}

In large classes of string compactifications the gauge group of the SM fields
includes an extra $U(1)_{B-L}$ gauge interaction. This is to be expected since
$U(1)_{B-L}$ is the unique flavour-independent $U(1)$ symmetry which is
anomaly free (in the presence of three right-handed neutrinos $\nu_R^a$,
needed also for the cancellation of mixed $U(1)_{B-L}$-gravitational anomalies).
In fact practically all semi-realistic MSSM-like string compactifications
constructed up to now have such an extra $U(1)_{B-L}$ interaction
and three right-handed neutrinos.

We will focus on this class of string compactifications
with an extra $U(1)_{B-L}$. Of course, such a gauge interaction forbids the
presence of Majorana masses for neutrinos, since they would violate
$U(1)_{B-L}$ gauge invariance. However, as pointed out in \cite{iu} (see also
\cite{bcw}),   string instanton effects may give rise to right-handed
neutrino Majorana masses under certain conditions. In particular
a crucial point is that the $U(1)_{B-L}$ gauge boson should get a Stuckelberg
mass from a $B\wedge F$ type of coupling. Here $B$ is a 2-index antisymmetric
field \footnote{ These tensors come e.g.  from the
RR-sector of Type II string theory. In $D=4$ they are
dual to pseudo-scalar fields $\eta_r$ which are the imaginary part
of complex scalar moduli fields, either complex structure
 moduli $U_r$
or Kahler moduli $T_r$, depending on the specific compactification.}
and $F$ is the $U(1)_{B-L}$
field strength. This mechanism is ubiquitous in string theory and
it plays an important role in $U(1)$ anomaly cancellation by the Green-Schwarz
mechanism (for a simple discussion see e.g. \cite{giiq}).
 Due to the presence of the $B\wedge F$ coupling, the
pseudo-scalar
$\eta$ (dual to the $B$ field) transforms under a  $U(1)_{B-L}$
gauge transformation of parameter $\Lambda (x)$ as:
\beq
\eta(x) \  \longrightarrow \ \  \eta(x) \ + \ q\ \Lambda (x)\;,
\eeq
with $q$ being some integer. The $\eta$ scalar has then a Higgs-like behavior
and gives a mass of order the string scale $M_s$ to the $U(1)_{B-L}$ gauge
boson. Thus, from the low-energy point of view the gauge symmetry
is just that of the SM (or possibly e.g. a $SU(5)$ extension).

As pointed out in \cite{iu,bcw}  in this class of models
string instantons can give rise to terms of the form
\beq
W_2\simeq e^{- {U_{ins}}}\,  {\nu_R\nu_R }\;,
\label{maj1}
\eeq
which give rise the right-handed  neutrino Majorana masses. Here
 $U_{ins}$ is a complex modulus scalar field characteristic
of the instanton and the particular compactification.
The point is that $Im\,U_{ins}$ is a linear combination of
axion-like fields including $\eta$ in such a way that under a
$U(1)_{B-L}$ gauge transformation transforms like
\beq
Im \, U_{ins}(x) \  \longrightarrow \ \  Im \,U_{ins}(x)  \ - \
2 \Lambda (x)  \;.
\eeq
Then the operator  $\exp(-U_{ins})$ has effective B-L charge=2 and
the operator (\ref{maj1}) is gauge invariant,
the gauge transformation of the neutrino bilinear is canceled by the
exponential.

This type of instanton contributions may
appear in all 4-dimensional string constructions but it is particularly 
intuitive in the case of Type IIA orientifolds with intersecting D6-branes \cite{bgkl,afiru,imr}
(see e.g. \cite{interev} for reviews and references).
These D6-branes have a 7-dimensional  worldvolume including Minkowski
space. The remaining 3-dimensions wrap a 3-cycle $\Pi$ of volume $V_{\Pi}$ in
the 6 compact dimensions.
In these models quarks and leptons appear as string excitations
localised  at D6-brane intersections. In the simplest configurations there are 4
different stacks of such D6-branes a,b,c,d associated to gauge groups
$U(3)_a\times SU(2)_b \times U(1)_c \times U(1)_d$. The $U(1)_{a,d}$
gauge symmetries correspond to baryon and lepton number respectively and $U(1)_c$
may be identified with the diagonal generator of right-handed weak isospin.
Out of these 3 $U(1)$'s only the linear combination $Y=Q_a/6-Q_c/2-Q_d/2$
corresponding to hypercharge remains light. The linear combination
$3Q_a+Q_d$ has triangle anomalies and gets a Stuckelberg mass as usual.
The remaining orthogonal linear combination $Y'=Q_a/6+Q_c/2-Q_d/2$ is anomaly-free
and again gets a Stuckelberg mass. The $U(1)_{B-L}$ generator is given by $Y+Y'$.

In these intersecting D6-brane models string instantons
\cite{ins1,ins2}
 are D2-branes with their
3-dimensional volume wrapping a 3-cycle  $\Pi_M$ on the 6 extra dimensions.
This is just like D6-branes, the main difference being that these D2-branes
are localised in $D=4$ space and time (that is why they are identified with instantons).
The action of these instantons is just the D2-brane classical action,
which is given by the Born-Infeld action which yields \footnote{In the present discussion
we will always work in string mass units with $M_s^2=(\alpha ')^{-1} =1$,
recovering the string mass dimensions for the final neutrino formulae.}
\beq
S_{D2}\, =\,
\frac
{ {V}_{ {\Pi}}}{  {\lambda } } +i \sum_r  {q_{M,r}}
 { \eta_r}\;,
\label{bi}
\eeq
where $V_{\Pi}$ is the 3-volume wrapped by the D2-brane, $\lambda $ is the
string dilaton and the imaginary piece is a linear combination with
integer coefficients of axion-like
RR-fields characteristic of the particular instanton $M$. For any given compactification and
instantons $S_{D2}$ may be written as a particular linear combination of
moduli fields $S_{D2}=U_{ins}$. In the particulñar case of Type IIA orientifolds
with intersecting D6-branes, they  are complex-structure moduli of the compact manifold.

%%%%%%%%%%%%%%%%%%%
\begin{figure}
\epsfysize=6cm
\begin{center}
\leavevmode
\epsffile{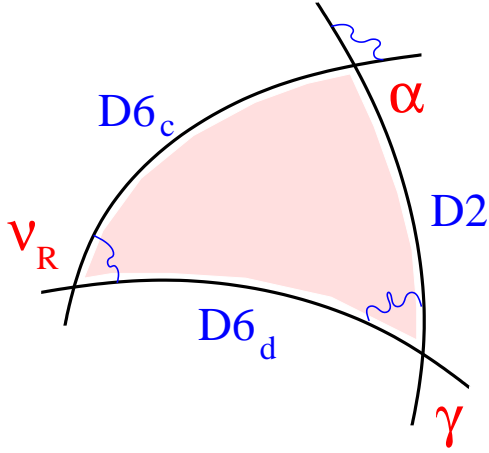}
\end{center}
\caption{World-sheet disk amplitude inducing a cubic coupling on the
D2-brane instanton action. The cubic coupling involves the
right-handed neutrinos lying at the intersection of the $c$ and $d$
D6-branes, and the  instanton fermion zero modes $\alpha $ and $\gamma$
from the D2-D6 intersections.}
\label{instanton3}
\end{figure}
%%%%%%%%%%%%%%%%%%%%%%%%%%%%

As described in \cite{iu,bcw,isu} there is in fact an extra contribution to the
instanton action which comes from possible intersections of the D2-instanton
and the relevant c and d D6-branes (see  figure 1 for a pictorial view). Right-handed
neutrinos come from string excitations around the intersections of d and c branes.
On the other hand the D2-instanton may intersect with the d and c branes and
at their intersections string excitations give rise to fermionic
zero modes $\alpha_i$ and $\gamma_i$. We will soon see that for a $\nu_R$ bilinear to be generated
the multiplicity of these modes must be two, i.e., $i,j=1,2$.
Note that, since D2-branes do not include
Minkowski space inside their volume,  $\alpha_i,\gamma_i$ are not $3+1$ dimensional
particles, like the $\nu_R$'s but rather 0+0 dimensional zero modes. They will
behave like Grassman variables over which one has to integrate. Indeed,
in computing the contribution of instantons to a given amplitude
(both with standard YM instantons and with string instantons of the type
here considered) one has to integrate over the moduli of the instanton and these
$\alpha_i,\gamma_i$ zero modes will be part of it.  Now, there are non-vanishing amplitudes
among the right-handed neutrinos $\nu_R^a$ and the zero modes which contribute to the
instanton action
\beq
 {S_{ins}}(\alpha,\gamma)\ = \  {d^{ij}_a}\ ( {\alpha_i}
  {\ \nu }  {^a \gamma_j}) \;  , \quad a=1,2,3  \ .
\eeq
Here $d_a^{ij}$ are coefficients which depend on the Kahler moduli of the compactification.
In order to obtain the induced superpotential one has to integrate
over the fermionic zero modes $\alpha_i,\gamma_i$ and
one obtains a superpotential coupling \footnote{There are additional factors of order one
coming from the quantum fluctuations of massive modes
(see e.g.\cite{bcw,abls}) . We set those terms to one in
the present analysis.} proportional to \cite{iu,bcw,isu}
\beq
\int d^2\theta
\int  {d^2\alpha\,  d^2\gamma }  \ e^{- {d^{ij}_a}\ (
{\alpha _i } {  \nu^a }
 {\gamma_j})}
\ = \  \int d^2\theta\, {\nu_a \nu_b}\,  (\, \epsilon_{ij}\epsilon_{kl}
 {d^{ik}_ad^{jl}_b}\, )\;  ,
\eeq
where we have made use of the Grassman integration rules $\int d\alpha=0$,
$\int d\alpha \alpha =1$ etc. Note that the fact that we have two $\alpha$
and $\gamma$ zero modes is crucial in order to obtain a bilinear.
This expression is multiplied by the exponential of
the classical action (\ref{bi}) so that the final expression
for the right-handed neutrino Majorana mass has the form
\beq
M^R_{ab}\ =\
{M_s}
(\,\epsilon _{ij} \epsilon_{kl}
 {d^{ik}_ad^{jl}_b}\, ) \exp(-U_{ins} ) \; , \quad a,b=1,2,3 \; ,
\eeq
where $M_s$ is the string scale and $\epsilon_{ij}$ is the 2-index antisymmetric unit tensor.
Note that the flavour information is encoded in the couplings $d^{ij}_a$. As discussed in detail
in \cite{isu}, the relevant D2-instantons have a gauge symmetry which is realised
only as a global symmetry in the effective $D=4$ spectrum. The simplest and most frequent situation
found up to now is that the global symmetry is $Sp(2)=SU(2)$, so that
the  $\alpha$ and $\gamma$ zero modes are doublets of $SU(2)$. In that situation
one can write $d^{ij}_a=d_a\epsilon^{ij}$ and the Majorana mass matrix takes a factorised form
\beq
M^R_{ab}\ =\
 2{M_s}
\sum_r d_a^{(r)}d_b^{(r)} \ e^{-U_r}\; ,
\label{sumins}
\eeq
where the sum goes over the different instantons which may contribute to this
Majorana mass term (in general there are several different instantons contributing).
As noted in \cite{isu}, this expression has an interesting flavour structure.
Indeed one can write
\beq
M^R \ \sim \ \sum_r  \
e^{-U_r}
\diag
(d_1^{(r)},d_2^{(r)},d_3^{(r)}) \cdot
\left(
\begin{array}{ccc}
1 & 1 & 1 \\
1 & 1 & 1 \\
1 & 1 & 1
\end{array}
\right)
\cdot
\diag(d_1^{(r)},d_2^{(r)},d_3^{(r)})  \;.
\label{matrilla}
\eeq
With this structure each instanton makes one
particular linear combination of $\nu_R$' s massive, leaving two linear combinations massless.
In particular one(two) instanton(s) contribution(s) would leave two(one) neutrinos
massless.
Thus with three or more contributing instantons generically all three get a mass.
Furthermore a hierarchy among the three eigenvalues may naturally appear taking
into account that each instanton will have in general a different
suppression factor $\exp(-Re\,U_{r})$. This will be one of the crucial ingredients of our
phenomenological analysis in the next sections.

Once (large) right-handed neutrino masses are generated the standard see-saw mechanism
\cite{seesaw} is expected to induce Majorana masses for the lightest eigenvalues
in the usual way, i.e. neutrino masses of the form
\beq
M_{ab}^L{\hbox{(see-saw}})\  =  \   \frac { \< {\overline H} \>^2}
{2M_s}
h_D^T (\sum_r \ d_a^{(r)}d_b^{(r)}  \  e^{-S_r})^{-1}\ h_D \;,
\eeq
where $h_D$  is the ordinary Yukawa coupling constant in $h_D^{ab}(\nu_R^a{\bar H}L^b)$.
The eigenvalues of these matrices are the ones which should be compared with
experiment.

%%%%%%%%%%%%%%%%%%%
\begin{figure}
\epsfysize=6cm
\begin{center}
\leavevmode
\epsffile{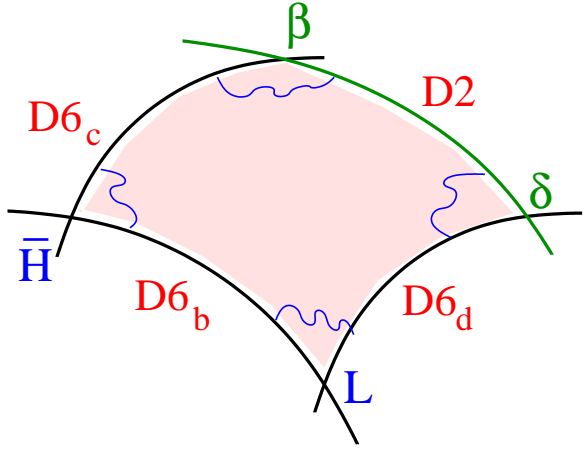}
\end{center}
\caption{World-sheet disk amplitude inducing a quartic  coupling on the
D2-brane instanton action. The  coupling involves the
Higgs ${\overline H}$ and left-handed leptons $L^a$ lying at the intersection of the $b$, $c$ and $d$
D6-branes, and the  instanton fermion zero modes $\beta $ and $\delta$
from the D2-D6 intersections.}
\label{instanton4}
\end{figure}
%%%%%%%%%%%%%%%%%%%%%%%%%%%%

As we mentioned, there is a second lepton number violating operator
which can be relevant for the structure of neutrino masses. This is the dim=5
Weinberg operator (in superpotential form)
\beq
 W_{W}\ =\   \frac {\lambda _{ab}}{M}(L^a{\overline H}L^b{\overline H})
\label{weinberg} \;.
\eeq
Once the Higgs fields get a vev, left-handed neutrino masses of
order $\< {\bar H} \>^2\lambda_{ab}/M$ are generated.
One important aspect of this operator is that it does not involve any
field beyond those of the SM (not even $\nu_R$' s) and does not directly involve
the see-saw mechanism.
String instantons can again give rise to such a superpotential in the class
of string models under consideration. A superpotential
of the form
\beq
 W_{W}\ =\  e^{-U_{ins}}\frac {\lambda _{ab} }{M_s} (L^a{\overline H}L^b{\overline H})
\label{weinberg} \;.
\eeq
may be generated in a way totally analogous to  the one discussed above for the
$\nu_R$ Majorana masses. The only main difference is that this time
the corresponding D2-instantons (which are different from the ones giving rise to
$\nu_R$ masses) have zero modes $\beta_i,\delta _i$ with quartic couplings
\beq
S_{ins}(\beta, \delta ) \ =  \ c^{ij}_a \ (\beta_i  (L^a {\overline H}) \delta_j)  \; .
\eeq
Again, in the simplest case with a $SU(2)$ symmetry operating in the $i,j$ indices
one has $c^{ij}_a=c_a\epsilon ^{ij}$ and a flavour factorised expression is
obtained for the left-handed neutrino Majorana masses
\beq
M_{ab}^L\  =  \   \frac { 2\< {\overline H} \>^2}
{M_s}
\sum_r \ c_a^{(r)}c_b^{(r)}  \  e^{-U_r} \;,
\label{sumons}
\eeq
where again the sum runs over possible different instantons contributing.
Note that the flavour structure of this mass matrix is
the same as that of Eq.~(\ref{matrilla}) so that a hierarchy of neutrino masses
naturally appears.

A comment is in order concerning the relationship between the see-saw mechanism
and the dim=5 Weinberg operator. For  constant field-independent
Majorana $\nu_R$ masses, the exchange of the $\nu_R$ fields gives rise to a
see-saw superpotential contribution to the Weinberg dim=5 term. On the other hand for
field dependent masses  like those generated from instantons, one cannot
write down the see-saw contribution in the form of a Weinberg
holomorphic superpotential.
So both contributions should be considered separately and in fact in the string models
different instantons contribute to both effects \cite{isu}.

In a given string compactification both mechanisms (see-saw and direct dim=5 operator)
may be present. Which is the dominant effect concerning the determination of the
masses and mixings of the observed neutrinos will be model dependent. In particular it will depend
on the particular values of the instanton actions $Re\,U_{ins}$, the value of the string scale $M_s$,
the values of the coefficients $c_{a},d_{a}$ and of the neutrino Yukawa couplings $h_D$.
In particular, if the $h_D$ couplings are small and there is little suppression from the
exponential factors $\exp(-U_r)$, the Weinberg operator might be dominant.
As we will see this case is particularly simple  because then one can directly
correlate the neutrino flavour structure to the string instanton mass generation formulae.
In the see-saw case the dependence on the string instanton effects is partially masked
by the dependence on the flavour dependent $h_D$ Yukawa couplings.

Looking at formulae (\ref{sumins}) and (\ref{sumons}) we see that the obtained neutrino masses
depend on three quantities, the string scale $M_s$, the instanton actions $Re\,U_r$ and the instanton
couplings  $c_a$ or $d_a$. Before entering into the phenomenological analysis
of the following sections let us review what can be said about these quantities.
Concerning the string scale $M_s$, it is in principle undetermined by present data
and may be as small as the TeV scale. On the other hand if we want to keep
gauge coupling unification and other simple features of MSSM-like scenarios,
identifying $M_s$ with the GUT-scale (i.e. of order $10^{16}$ GeV) is
an attractive option. The Weinberg operator may induce neutrino masses of order
$10^{-1}-10^{-2}$ eV  as long as
\beq
M_s\ < \ 10^{15} (c_a)^2 e^{-Re\,U_W} \:\mbox{GeV}\;.
\eeq
If this was the dominant source of neutrino masses, this would seem
to favor values of the string scale below the unification
scale $10^{16}$ GeV. However, if there are a number of different instantons
contributing and $Re\,U_W$ is small, it could still be computable with
$M_s$ of order the unification scale.

If the dominant source of observed neutrino masses were the see-saw
mechanism, one can obtain neutrino masses of order
$10^{-1}-10^{-2}$ eV  as long as
\beq
M_s\ < \ 10^{15} \frac {h^2}{d_a^2 } e^{Re\,U_M} \:\mbox{GeV}\;,
\eeq
where $h$ is the size of the largest neutrino Yukawa coupling. In this case
the size of the string scale is essentially unconstrained.

Concerning the values of the string instanton actions $Re\,U_r$, it is important to remark that,
unlike the standard YM instantons of electroweak interactions, string instanton effects
are not particularly suppressed, since $Re\,U_r$ are unrelated to the (inverse) gauge couplings of any
SM interactions. This means that the exponential factors $\exp(-Re\,U_r)$ appearing in the
amplitudes may be in fact of order one or, say, $O(1/10)$ and hence the neutrino operators
we are talking about need not be particularly suppressed.
The actions of each individual instanton, proportional to $Re\,U_r$, are generically
different. For example, in section 6.4 of \cite{isu} an example of compactification
is shown in which there are three  instantons contributing to
right-handed neutrino masses whose actions
are on the ratios $1:3.8:16.2$. The overall normalization depends on the value of the
Type II string dilaton, which is a free parameter in perturbative compactifications.
This example illustrates how indeed a hierarchy of neutrino mass eigenvalues
is possible in the present context.

Concerning the amplitudes $c_a$ and $d_a$, $a=1,2,3$, they are obtained from string correlators
involving the  chiral field operators ${\bar H}L^a$ and $\nu_R^a$ respectively and the
fermionic instanton zero modes $\beta_i,\delta_i$ and $\alpha_i,\gamma_i$, $i=1,2$.
In the case of intersecting D6-brane models they are in general
functions of the Kahler moduli $T_k$
of the compactification.
 If the string compactification
involves a known conformal field theory (CFT) like in toroidal (or orbifold)
models or models obtained from  Gepner orientifolds \cite{schell} (see also \cite{isu} and references therein),
$c_a$,$d_a$ are computable in principle. In practice only for toroidal models such
computations are available at the moment (very much like it happens with ordinary
Yukawa couplings). However, although there have been constructed  Type IIA
orbifold orientifolds with intersecting D6-branes and a MSSM-like spectrum (see \cite{interev}
for reviews and references)  none of them have a massive $U(1)_{B-L}$ as required
for the present instanton mechanism to work. On the other hand there are
non supersymmetric intersecting D6-brane models in which such massive $U(1)_{B-L}$
gauge bosons occur. As pointed out in \cite{iu}, for such models
the $d_a(T_k)$ amplitudes are analogous to those of ordinary Yukawa
couplings \cite{cim}
and they are typically proportional to (products of) Jacobi theta functions
$\theta[\delta^a ,0]( \phi_i,  T_k )$. Here $\delta ^a$ are some fractional
numbers which depend on the generation number $a=1,2,3$, $T_k$ are Kahler moduli and  $\phi_I $ are scalar moduli
fields which parameterise the location of the D-branes in extra dimensions.
However, the non-SUSY toroidal examples  discussed in \cite{iu} need to be completed
since they require the presence of further backgrounds in order to
get the adequate number of instanton zero modes for the neutrino mass operators
to be generated. Still this at least illustrates what type of functions
could appear in more realistic computations. For example, we will see
in the phenomenological applications in the next section that, e.g.,
in order to have a small $\theta_{13}$ in the neutrino mixing matrix,
a suppressed  $c_1$ amplitude for the leading instanton would be required.  So we might want to impose
such a condition on candidate string compactifications. In this connection it is perhaps
worth noticing that Jacobi theta functions do
 vanish in  particular symmetric points.

Given our discussion above, our approach in the present paper concerning the
amplitudes $c_a$, $d_a$ will be mostly phenomenological. We will address ourselves the
question: under what circumstances are the present class of instanton induced
neutrino masses consistent with present experimental constraints?
In the next section we will see that under very mild constraints on the $c_a$
coefficients the instanton generated Weinberg operator will be consistent
with present experimental data on neutrino masses and mixings.
Furthermore we will see that if the $c_a$ amplitudes go along certain
directions in flavour space, e.g.,  tri-bimaximal neutrino mixing
may be obtained.

\section{Neutrino masses and mixing angles}
In general, the leptonic mixing matrix (PMNS matrix), is given in the standard PDG parameterisation as
\begin{eqnarray}\label{Eq:StandardParametrization}
 U_{\mathrm{PMNS}} = \left(
  \begin{array}{ccc}
  c_{12}c_{13} &
  s_{12}c_{13} & s_{13}e^{-i \delta}\\
  -c_{23}s_{12}-s_{13}s_{23}c_{12}e^{i \delta} &
  c_{23}c_{12}-s_{13}s_{23}s_{12}e^{i \delta}  &
  s_{23}c_{13}\\
  s_{23}s_{12}-s_{13}c_{23}c_{12}e^{i \delta} &
  -s_{23}c_{12}-s_{13}c_{23}s_{12}e^{i \delta} &
  c_{23}c_{13}
  \end{array}
  \right) \, P_\mathrm{Maj} \; ,
\end{eqnarray}
where we have used the abbreviations $s_{ij} = \sin(\theta_{ij})$ and $c_{ij} = \cos(\theta_{ij})$. 
Here $\delta$ is the so-called Dirac CP violating phase which is in
principle measurable in neutrino oscillation experiments, and
$P_\mathrm{Maj} = \mathrm{diag}(e^{i \tfrac{\alpha_1}{2}},
e^{i \tfrac{\alpha_2}{2}}, 0)$ contains the Majorana phases $\alpha_1,
\alpha_2$. The PMNS mixing receives
 contributions from the matrix $V_{e_\mathrm{L}}$ diagonalizing the mass matrix of
the charged leptons and from $V_{\nu_\mathrm{L}}$ diagonalizing the neutrino mass matrix,
\begin{eqnarray}\label{Eq:PMNS_Definition}
U_{\mathrm{PMNS}}
= V_{e_\mathrm{L}} V^\dagger_{\nu_\mathrm{L}} \,.
\end{eqnarray}
In the following, we assume that the large mixing in the lepton sector originates
from the neutrino mass matrix. In fact such large mixings are
generically expected in the present context, given  the very different origins of
the neutrino masses from string instantons, compared to the masses for  charged leptons
 and quarks.
Under this assumption, we may treat the small mixings of the charged lepton mass matrix as
a perturbation.

\subsection{Masses and mixings from the Weinberg operator}
We first discuss the case that the contribution from the
 dimension 5 Weinberg operator dominates the neutrino mass matrix. As we said, in this case the
left-handed neutrino mass matrix is directly related to the instanton contribution discussed
in the previous section.

\subsubsection{The general normal hierarchy case}

When the Weinberg operator dominates,
the instanton-induced neutrino mass matrix can be written in the form
\begin{eqnarray}\label{Eq:WeinbergOpDom_1}
 M_{ij}^L = \sum_r I_r c_i^{(r)}c_j^{(r)},
\label{masafac}
\end{eqnarray}
where we have defined
\begin{eqnarray}
 I_{r}= \frac{\< {\overline H} \>^2}{M_s}2e^{-S_r}\;.
\end{eqnarray}
Let us consider first the scenario in which we have three instanton contributions to the
neutrino mass matrix ($r=1,2,3$). Let us assume there is some
 (eventually mild) hierarchy, in particular
\begin{eqnarray}\label{SDcondition}
|I_3 c_i^{(3)} c_j^{(3)} |\gg |I_2 c_i^{(2)} c_j^{(2)}| \gg |I_1 c_i^{(1)} c_j^{(1)}|\;.
\end{eqnarray}
This may be motivated by a hierarchy of the instanton factors,
\begin{eqnarray}
e^{-Re(S_3)} \gg e^{-Re(S_2)}\gg e^{-Re(S_1)}\;,
\end{eqnarray}
As we mentioned in the previous section, such modest hierarchies are likely in
 orientifold compactifications (see e.g.\cite{isu}).
 In this situation  we can extract analytically the conditions under
which the generated neutrino masses and mixing angles are
compatible with the experimental results.

In order to simplify the following discussion of the leptonic mixing angles and CP
phases resulting from neutrino masses of the form in  Eq.~(\ref{Eq:WeinbergOpDom_1}), we define
\begin{eqnarray}
\overline{c}_i^{(3)}=\sqrt{I_3}c_i^{(3)}\;, \quad
\overline{c}_i^{(2)}=\sqrt{I_2 }c_i^{(2)}\;,\quad
\overline{c}_i^{(1)}=\sqrt{I_1 }c_i^{(1)}\;,
\end{eqnarray}
and furthermore
\begin{eqnarray}
\phi_{i}^{(3)} = \arg (c_i^{(3)})\;, \quad
\phi_{i}^{(2)} = \arg (c_i^{(2)})\;, \quad
\phi_{i}^{(1)} = \arg (c_i^{(1)})\;,
\end{eqnarray}
for $i = 1,2,3$. In the limit of Eq.~(\ref{SDcondition}), and using that the observed mixing angle $\theta_{13}$ is small, the mixing angles of the PMNS matrix (using the standard PDG parameterisation \cite{PDG}) are then given as
\begin{eqnarray}
\tan\theta_{23}\approx\frac{|\overline{c}_2^{(3)}|}{|\overline{c}_3^{(3)}|}
\end{eqnarray}
\begin{eqnarray}
\tan\theta_{12}\approx\frac{|{\overline{c}}_1^{(2)}|}{{c}_{23}|
\overline{c}_2^{(2)}|\cos{\tilde{\phi_2}}-s_{23}|\overline{c}_3^{(2)}|\cos{\tilde{\phi_3}}}
\end{eqnarray}
\begin{eqnarray}
\theta_{13}\approx e^{i(\tilde\phi +
\phi_{1}^{(2)}-\phi_{2}^{(3)})}\frac{|\overline{c}_1^{(2)}|
(\overline{c}_2^{(3)*}\overline{c}_2^{(2)}+\overline{c}_3^{(3)*}\overline{c}_3^{(2)})}{[|\overline{c}_2^{(3)}|^2+
|\overline{c}_3^{(3)}|^2]^{\frac{3}{2}}} + e^{i(\tilde\phi + \phi_{1}^{(3)}-\phi_{2}^{(3)})}
 \frac{|\overline{c}_1^{(3)}|}{\sqrt{|\overline{c}_2^{(3)}|^2+|\overline{c}_3^{(3)}|^2}}\;,
\end{eqnarray}
where we have defined
\begin{eqnarray}
\tilde{\phi}_2 = \phi_{2}^{(2)} - \phi_{1}^{(2)}- \tilde{\phi} + \delta\;,
\end{eqnarray}
\begin{eqnarray}
\tilde{\phi}_3 = \phi_{3}^{(2)} - \phi_{2}^{(2)} +
\phi_{2}^{(3)}- \phi_{3}^{(3)}-\tilde{\phi} + \delta\;.
\end{eqnarray}
$\delta$ and $\tilde{\phi}$ are determined from the condition/convention that $\tan\theta_{12}$ and $\theta_{13}$ are real and positive, respectively.
Under the above conditions, the neutrino masses are given by
\begin{eqnarray}
m_3 = (|\overline{c}_2^{(3)}|^2 + |\overline{c}_3^{(3)}|^2)\;,
\end{eqnarray}
\begin{eqnarray}
m_2 = \frac{|\overline{c}_1^{(2)}|^2}{s_{12}}\;,
\end{eqnarray}
\begin{eqnarray}
m_1 = \textit{O}(|\overline{c}^{(1)}|^2)\;.
\end{eqnarray}

We note that from a technical point of view, the procedure which has been used for
extracting the neutrino parameters is equivalent to the  one for see-saw models of
neutrino masses with sequential right-handed neutrino dominance \cite{King:1998jw}. However, it
is applied here in a different physical context, namely that of  neutrino masses from string
theory instantons which generate the Weinberg operator.

\begin{table}
\begin{center}
\begin{tabular}{l|ccc}
&$\vphantom{\sqrt{\big|}}$ Best-fit value & Range & C.L. \\
  \hline
$\vphantom{\sqrt{\big|}}$\(\theta_{12}\) [\({\:}^\circ\)]& \(33.2\) & \(29.3 - 39.2\) &
\(99 \%
\;(3\sigma)\)\\
$\vphantom{\sqrt{\big|}}$\(\theta_{23}\) [\({\:}^\circ\)]& \(45.0\) & \(35.7 - 55.6\) &
\(99 \%
\;(3 \sigma)\)\\
$\vphantom{\sqrt{\big|}}$\(\theta_{13}\) [\({\:}^\circ\)]& \(-\) &
\(0.0-  11.5\) &     \(99 \% \;(3 \sigma)\)\\
$\vphantom{\sqrt{\big|}}$\(\Delta m^2_{21}\) [eV\(^2\)] & \(7.9 \cdot 10^{-5}\)&
\(7.1\cdot 10^{-5} - 8.9\cdot 10^{-5}\) & \(99 \%
\;(3 \sigma)\) \\
$\vphantom{\sqrt{\big|}}$\(|\Delta m^2_{31}|\) [eV\(^2\)] & \(2.6 \cdot 10^{-3}\)&
\(2.0\cdot 10^{-3} - 3.2\cdot 10^{-3}\) & \(99 \%
\;(3 \sigma)\)\hfill
\end{tabular}
\caption{
Experimental results for the neutrino mixing angles and mass squared differences, taken from the recent global fit of Ref.~\cite{Maltoni:2004ei} to the present neutrino oscillation data.
}\label{tab:ExpData}
\vspace{-0.35cm}
\end{center}
\end{table}

Let us now turn to the conditions for consistency with experiment.
The present experimental status is summarised in Tab.~\ref{tab:ExpData}. We see that under the ``sequential dominance'' assumptions of Eq.~(\ref{SDcondition}) the following general conditions are imposed on the
parameters $c_i^{(r)}$:
\begin{itemize}
\item Large, nearly maximal, mixing $\theta_{23} \approx \pi/4$ implies that
$| c_2^{(3)}|\simeq |c_3^{(3)}|$.
\item Large (but non-maximal) $\theta_{12}$ implies that $|c_1^{(2)}|\simeq |c_2^{(2)}|\simeq |c_3^{(2)}|$, or at least that $|c_1^{(2)}|$ and either $|c_2^{(2)}|$ or $|c_3^{(2)}|$ are of the same order.
\item Small $\theta_{13}$ requires that $|c_1^{(3)}|/|c_3^{(3)}|$ is small.
\end{itemize}
Generically, coefficients $c_i^{(r)}$ of ${\cal O}(1)$ are a typical expectation in the present scheme.
This means, large mixings  are not only
easy to accommodate, but are even expected in the considered scenario.
However, the explicit values depend on the details of the model,
 and small (or even vanishing) values for the $c_a^{(r)}$ amplitudes
 may emerge in particular examples.
The condition $|c_1^{(3)}| \ll |c_3^{(3)}|$ may thus give us information/constraints
on which string constructions may be fully successful in describing the neutrino data.

For a hierarchical neutrino spectrum, the conditions of Eq.~(\ref{SDcondition}) imply that the
particular parameters $c_i^{(1)}$  (corresponding to the most suppressed
instanton effect) do not play a significant role for the mixing angles.
In fact, only two instantons are required to give masses $m_2$ and $m_3$
 to two linear combination of neutrinos fields, while one of the neutrinos could remain massless.
The remaining constraint is that the two instanton contributions proportional to $e^{-S_2}$
and $e^{-S_3}$ have to generate neutrino masses $m_2 \approx \sqrt{\Delta m^2_{21}}$ and
 $m_3 \approx \sqrt{|\Delta m^2_{31}|}$.

\subsubsection{Normal hierarchy and tri-bimaximal neutrino mixing}
One of the most popular proposed structures for neutrino mixing is that of tri-bimaximal mixing \cite{tribi}.
We would like now to study under what   conditions the neutrino mass matrix
 from string theory instantons via the Weinberg operator, i.e.\ of the form of Eq.~(\ref{Eq:WeinbergOpDom_1}),
 can give rise to tri-bimaximal lepton mixing.
Tri-bimaximal lepton mixing is a pattern of neutrino mixing angles postulated by \cite{tribi}, where the PMNS matrix is given by
\begin{eqnarray}
U_\mathrm{tri} = \left(\begin{array}{ccc}
\!\sqrt{2/3}&1/\sqrt{3} &0\!\\
\!-1/\sqrt{6}&1/\sqrt{3}&-1/\sqrt{2}\!\\
\!-1/\sqrt{6}&1/\sqrt{3}&1/\sqrt{2}\!
\end{array}\right).
\label{bitritri}
\end{eqnarray}
In the standard PDG parameterisation \cite{PDG}, this corresponds to $\theta_{12} = \arcsin (1/\sqrt{3})
 \approx 35.3^\circ, \theta_{13}=0$ and $\theta_{23}= \arcsin (1/\sqrt{2})=45^\circ$ in the lepton sector.
 The PMNS matrix is usually given in
the basis where the so-called ``unphysical phases'' are eliminated by absorbing a global
 phase factor in the definition of the lepton doublets. Since in our case neutrinos have
 Majorana masses, the PMNS matrix is multiplied by an additional phase matrix $P_\mathrm{Maj}
 = \mathrm{diag}(e^{i \tfrac{\alpha_1}{2}},
e^{i \tfrac{\alpha_2}{2}}, 0)$ from the right, which contains the Majorana phases $\alpha_1,\alpha_2$.
As stated earlier, we assume that the large mixing in the lepton sector originates
from the neutrino mass matrix, such that we may treat the small mixings of the
charged lepton mass matrix as a perturbation. We will consider the minimal case of two instantons,
the minimal number required in order generate two massive neutrinos.

Let us try to find an example for tri-bimaximal mixing of
the neutrino mass matrix from instantons.
%The simplest case seems
%to be one in which the string theory instantons directly generate the Weinberg operator,
% and that this provides the dominant contribution to neutrino masses.
% In this case, the neutrino mass matrix is given by Eq.~(\ref{Eq:MabLfromInstantons}). Defining the neutrino
%masses as
%\begin{eqnarray}
%I_r = \frac{\< \bar H \>}{M_s} \,2\, e^{-S_r}\; ,
%\end{eqnarray}
%where $r = 1,2,3$, Eq.~(\ref{Eq:MabLfromInstantons}) can be rewritten as
%\begin{eqnarray}\label{Eq:MabLfromInstantons2}
%M^L_{ab} = \sum_r I_r c_a^{(r)} c_b^{(r)} \;,
%\end{eqnarray}
%%where we have normalised the $c^{(r)}$ such that $m_r$ are the eigenvalues, i.e.\
%the neutrino masses.
%$a,b = 1,2,3$ are family indices. At least two instantons are required
%in order to generate two massive neutrinos, as required by the experimental data.
To start with, we may assume a normal hierarchy for the neutrino masses,
i.e.\ $m_1 \ll m_2 \approx \tfrac{1}{5} m_3$, and set $m_1$ to zero.
Using the expression (\ref{bitritri}) one obtains for the
the neutrino mass matrix with tri-bimaximal mixing
\begin{eqnarray}\label{Eq:Mtribi}
M^\mathrm{tri} &=& U_\mathrm{tri} \, \mbox{diag}(0,m_2 \, e^{i \alpha_2},m_3)\, U^T_\mathrm{tri}
\nonumber \\
&=&
\frac{m_2\, e^{i \alpha_2}}{3} \begin{pmatrix}
 1 &  1 & 1 \\
 1 &  1 & 1 \\
 1 &  1&  1 \\
\end{pmatrix}
+
\frac{m_3}{2} \begin{pmatrix}
 0 &  0 &  0 \\
 0 &  1 & -1 \\
 0 & -1 &  1 \\
\end{pmatrix}.
\end{eqnarray}
By comparing this form with Eq.~(\ref{masafac}), we see that a possibility to obtain this structure is to identify
$I_2 = m_2, I_3 = m_3$ and to choose the coefficients $c_i^{(2)},c_i^{(3)}$ as
\begin{eqnarray}
(c_1^{(2)},c_2^{(2)},c_3^{(2)}) &=& \frac{1}{\sqrt{3}}\, e^{i \tfrac{\alpha_2}{2}}\,(1,1,1)  \; ,\\
(c_1^{(3)},c_2^{(3)},c_3^{(3)}) &=& \frac{1}{\sqrt{2}}(0,-1,1) \; .
\end{eqnarray}
We would like to remark that this is one particular possibility, not the most general case. However, with a hierarchy among the neutrino masses and a hierarchy among the instanton contributions, it is suggestive that one instanton generates $m_2$ and the other one $m_3$.
Note also that the ``normalisation'' of the ``flavour vectors'' $c_i^{(r)}$ can
be changed to $c^{(r)} \to {\cal N} c^{(r)}$ by
choosing $ I_r = {\cal N}^{-2} m_r$ instead of  $I_r = m_r$.

We see that obtaining precisely the structure of tri-bimaximal mixing requires the
flavour vectors $c_a^{(2)},c_a^{(3)}$ to align  along specific flavour directions
\footnote{Note in particular that in order to exactly reproduce the tri-bimaximal mixing
matrix there should be three instantons with flavour vectors $c_a^{(r)}$ aligning
along the Cartan subalgebra generators of a $U(3)$ group.}.
 On the other hand
obtaining masses and mixings compatible with experiment require much milder
constraints on
the flavour vectors, as we discussed in the previous sections.

\subsubsection{The inverse hierarchy case}
Experimentally, two possibilities for the ordering of the neutrino masses are allowed:
The so-called normal ordering where $m_1 < m_2 < m_3$, and the so-called
inverse ordering where $m_3 < m_1 < m_2$. If in the latter case $m_3 \ll m_1 \lesssim m_2$,
 the neutrino spectrum is called inverse hierarchical. String theory instantons
 can in principle also give rise to
this scenario. However, we have to keep in mind that the splitting between $m_1$ and $m_2$ is very small,
 and that it would have to be explained why $m_1 \approx m_2$.
 Within the string instanton point of view, this would require
the presence of two instantons $D2_1$, $D2_2$  with  approximately the same actions $S^{(1)}$,$S^{(2)}$
but with very different flavour vectors $c_a^{(1)},c_a^{(2)}$.
Since the action is given essentially by the size of the wrapped 3-volume in extra dimensions
and the latter are expected to be generically different, a certain amount of fine-tuning
would be required. Different values for the different actions $S^{(r)}$, typically
leading to some hierarchy seems more generic. Aiming at
completeness, we will nevertheless consider the inverse hierarchy case as well.

Examples of patterns for the relevant coefficients $c_i^{(1)}$ and $c_i^{(2)}$ in this case can be found easily
 following the strategy used in the above subsection. Since general analytic formulae are rather lengthy, we will
focus on  a particular example here, noting that many variations and alternative patterns are possible and allowed by
the experimental data. As above, we consider the example of tri-bimaximal mixing since approximate tri-bimaximality is well compatible with the present experimental data.
For the inverse hierarchy case, in principle tri-bimaximal mixing (and other patterns of neutrino mixing angles compatible with experiment) could be realised as well.
Setting $m_3 = 0$, the neutrino mass matrix with tri-bimaximal mixing has the form
\begin{eqnarray}
M^\mathrm{tri} &=& U_\mathrm{tri} \, \mbox{diag}(m_1\, e^{i \alpha_1},m_2\,
e^{i \alpha_2},0)\, U^T_\mathrm{tri} \nonumber \\
&=&
\frac{m_1 \, e^{i \alpha_1}}{6} \begin{pmatrix}
 4  & -2 & -2 \\
 -2 &  1 &  1 \\
 -2 &  1 &  1 \\
\end{pmatrix}
+
\frac{m_2 \, e^{i \alpha_2}}{3} \begin{pmatrix}
 1 &  1 & 1 \\
 1 &  1 & 1 \\
 1 &  1&  1 \\
\end{pmatrix},
\end{eqnarray}
and only two string instantons are required (in the most minimal case) to give neutrino
masses $m_1$ and $m_2$.
Again, comparing with Eq.~(\ref{masafac}) we see that a possible choice is $I_1 = m_1, I_2 = m_2$
(with $I_1\simeq I_2$) and
\begin{eqnarray}
(c_1^{(1)},c_2^{(1)},c_3^{(1)}) &=& \frac{1}{\sqrt{6}}\, e^{i \tfrac{\alpha_1}{2}}\,(-2,1,1)  \; ,\\
(c_1^{(2)},c_2^{(2)},c_3^{(2)}) &=& \frac{1}{\sqrt{3}}\, e^{i \tfrac{\alpha_2}{2}}\,(1,1,1) \; .
\end{eqnarray}

\subsubsection{Quasi-degenerate neutrino spectrum}
We now turn to the general case, which includes the cases
of quasi-degenerate (or partially-degenerate) neutrino masses with $m_1,m_2,m_3$ non-zero and with typically two of them being nearly degenerate in mass.
In our scheme  one  can in principle accommodate  this scenario as well.
However, now the splitting between $m_1, m_2$ and $m_3$ are very small, and this almost degeneracy
of the mass eigenvalues would have to be explained. Explicitly, the masses have
 to satisfy the experimental constraints, i.e.\ $m_2^2 - m_1^2 \approx 7.9 \times 10^{-5}$ eV$^2$,
$|m_3^2 - m_1^2| \approx 2.6 \times 10^{-3}$ eV$^2$ (c.f.\ Tab.~\ref{tab:ExpData}), while $m_1 \approx m_2 \approx m_3$ are
much larger than the mass splitting.
From the string instanton  point of view, this would require again having
three different instanton with almost identical action but very different
flavour vectors. Although possible such situation would require some
fine tuning and is generically unexpected.

In order to give an example for a pattern of $c_i^{(r)}$ compatible with the experimentally found mixing angles, let us consider again the concrete example of tri-bimaximal mixing. Tri-bimaximal mixing for quasi-degenerate neutrinos could be realised with three instantons with $c^{(1)},c^{(2)},c^{(3)}$, chosen as
\begin{eqnarray}
(c_1^{(1)},c_2^{(1)},c_3^{(1)}) &=& \frac{1}{\sqrt{6}}\, e^{i \tfrac{\alpha_1}{2}}\,(-2,1,1)  \; ,\\
(c_1^{(2)},c_2^{(2)},c_3^{(2)}) &=& \frac{1}{\sqrt{3}}\, e^{i \tfrac{\alpha_2}{2}}\,(1,1,1) \; ,\\
(c_1^{(3)},c_2^{(3)},c_3^{(3)}) &=& \frac{1}{\sqrt{2}}(0,-1,1)\;.
\end{eqnarray}
and with $I_1 = m_1, I_2 = m_2, I_3 = m_3$ (and $I_1\simeq I_2\simeq I_3$)..

\subsection{The see-saw case}

%Here discuss the see-saw case. It is less straightforward.
%We will assume a diagonal Dirac neutrino mass structure...
%(I dont have the file with diagonal matrix different from
%the identity...)

Up to now we have considered the case in which the leading contribution to
the observed neutrino masses comes from the Weinberg operator. Let us consider now
the inverse case in which the see-saw mechanism gives the leading contribution to
neutrino masses.
In general, the see-saw contribution to the neutrino mass matrix can depend significantly
on the structure of the neutrino Yukawa matrix $h_D$, leading to a large variety
of possible patterns of $h_D$ and $M^R_{ab}$ consistent with
the experimental neutrino data (assuming  again small charged lepton mixing, as before).

Obtaining analytic formulae for the most general see-saw case is difficult.
In the following, we discuss the special case
in which
 only small mixing stems from the neutrino Yukawa matrix $h_D$.
Explicitly, we will consider the limit that $h_D$ is diagonal, i.e.
\begin{eqnarray}
h_D = \mbox{diag}({y_e^{(\nu)}},{y_\mu^{(\nu)}},{y_\tau^{(\nu)}}) \; .
\end{eqnarray}
Small mixing induced by $h_D$ may be treated as a perturbation and can be included in a straightforward way.
We will furthermore assume that the see-saw contribution dominates the neutrino mass matrix.

The neutrino mass matrix is then given by
\begin{eqnarray}
M^L_{ab} =  \left(\sum_r \frac{d_a^{(r)}}{(h_D^T)_{aa}} \frac{d_b^{(r)}}{(h_D)_{bb}} \widetilde{I}_r^{-1}\right)^{-1} \;,
\end{eqnarray}
where we have introduced
\begin{eqnarray}
\widetilde{I}_r = \frac{\< \bar H \>^2}{2 M_s} \frac{1}{e^{-S_r}}\;.
\end{eqnarray}
and which defines the quantity
\begin{eqnarray}
\widetilde M^L_{ab} := (M^L_{ab})^{-1} = \sum_r \frac{d_a^{(r)}}{(h_D^T)_{aa}} \frac{d_b^{(r)}}{(h_D)_{bb}} \widetilde{I}_r^{-1}\;.
\end{eqnarray}
The indices of the matrix $\widetilde M^L_{ab}$ must be understood  as those coming from the numbers $d_a^{(r)}$ and
$d_b^{(r)}$, while those of the matrix $M^L_{ab}$ must be calculated as in the usual matrix calculus.

As an example, we now discuss how to choose the coefficients $d_a^{(r)}$ in order to realise tri-bimaximal neutrino mixing. We note that this procedure can be readily generalised to any other desired pattern of neutrino mixings.
We first observe that if we find $d_a^{(r)}$ such that $U_\mathrm{tri}$ diagonalises $\widetilde M^L_{ab}$, then also $M^L_{ab}$ has tri-bimaximal form,
\begin{eqnarray}
U^T_\mathrm{tri}\,\widetilde M^L \,U_\mathrm{tri} = \mbox{diag}(\tfrac{1}{m_1},\tfrac{1}{m_2},\tfrac{1}{m_3})\; \Rightarrow \;
U^T_\mathrm{tri}\,(\widetilde M^L)^{-1} \,U_\mathrm{tri} = \mbox{diag}(m_1,m_2,m_3)\;,
\end{eqnarray}
since $U_\mathrm{tri}$ is orthogonal.
The form of $\widetilde M^L$ required to realise tri-bimaximal mixing is thus given by
\begin{eqnarray}
\widetilde M^L &=& U_\mathrm{tri} \,\mbox{diag}(\tfrac{1}{m_1\, e^{i \alpha_1}},\tfrac{1}{m_2\, e^{i \alpha_2}},\tfrac{1}{m_3})\,U^T_\mathrm{tri}
 \\&&
\frac{1}{6\, m_1\,e^{i \alpha_1}}
\begin{pmatrix}
 4  & -2 & -2 \\
 -2 &  1 &  1 \\
 -2 &  1 &  1 \\
\end{pmatrix}
+
\frac{1}{3\,m_2\, e^{i \alpha_2}}
\begin{pmatrix}
 1 &  1 & 1 \\
 1 &  1 & 1 \\
 1 &  1&  1 \\
\end{pmatrix}
+
\frac{1}{2\,m_3 }
\begin{pmatrix}
 0 &  0 &  0 \\
 0 &  1 & -1 \\
 0 & -1 &  1 \\
\end{pmatrix}.  \nonumber
\end{eqnarray}
A possible choice for the $d_a^{(r)}$ is therefore (analogous to the Weinberg operator cases)
$\widetilde I_1 = m_1,\widetilde I_2 = m_2,\widetilde I_3 = m_3$ and
\begin{eqnarray}\label{Eq:CondSeesawDiaghD}
\left(\frac{d_1^{(1)}}{{y_e^{(\nu)}}}, \frac{d_2^{(1)}}{{y_\mu^{(\nu)}}},\frac{d_3^{(1)}}{{y_\tau^{(\nu)}}}\right) &=& \frac{1}{\sqrt{6}}\, e^{-i \tfrac{\alpha_1}{2}}\,(-2,1,1)  \; ,\\
\left(\frac{d_1^{(2)}}{{y_e^{(\nu)}}}, \frac{d_2^{(2)}}{{y_\mu^{(\nu)}}},\frac{d_3^{(2)}}{{y_\tau^{(\nu)}}}\right) &=& \frac{1}{\sqrt{3}}\, e^{-i \tfrac{\alpha_2}{2}}\,(1,1,1) \; ,\\
\left(\frac{d_1^{(3)}}{{y_e^{(\nu)}}}, \frac{d_2^{(3)}}{{y_\mu^{(\nu)}}},\frac{d_3^{(3)}}{{y_\tau^{(\nu)}}}\right) &=& \frac{1}{\sqrt{2}}(0,-1,1)\;.
\end{eqnarray}
As discussed for the Weinberg operator case, only two of the right-handed neutrinos  are
relevant in the limit of the normal and inverse hierarchy cases.

We see that if $(h_D)_{aa} = {y_a^{(\nu)}}$, $a=e,\mu,\tau$, are hierarchical,
then also the $d_a^{(r)}$ (for all $r$) would have to have a very similar hierarchical structure,
in order to generate large neutrino mixing. Although  possible in principle,
 this would be  a significant constraint on models. On the other hand, with $(h_D)_{aa} = {y_a^{(\nu)}}$ being all of
the
same order, the conditions of Eq.~(\ref{Eq:CondSeesawDiaghD}) could be comparatively easier to satisfy and
large neutrino mixing angles would be a generic expectation.

A different possibility would of course be that only small mixing is induced by $M^R$ and that large mixing originates from $h_D$. In this case, we recover the known conditions on $h_D$ and $M^R$ discussed extensively in the literature on conventional see-saw models  \cite{ModelReviews}.

\subsection{The general case: Weinberg operator and See-saw}

More generally, instantons may generate the Weinberg operator for neutrino masses, which provides a direct mass term
for (some of)
the three light neutrinos after EW symmetry breaking, as well as the Majorana mass matrix for the right-handed neutrinos. The full neutrino mass matrix $M$ has dimension
$6 \times 6$,
\begin{eqnarray}\label{Eq:GeneralMatrix}
M =
\begin{pmatrix}
M^L     & v h_D^T \\
v h_D & M^R
\end{pmatrix}.
\end{eqnarray}
Beyond the discussion of the previous sections, there is the possibility that the contributions from the see-saw mechanism and from the Weinberg operator both contribute with similar strength to the mass matrix of the light neutrinos. For example, one may have the case that one of the contributions generates the dominant term in Eq.~(\ref{Eq:Mtribi}), while the other generates the sub-dominant one.

Finally, it is also possible in principle that some of the right-handed neutrinos
could obtain rather small masses, such that there are more than three light neutrino mass eigenstates (or
right-handed neutrinos close to the EW scale).
In specific string  models, all ingredients of the neutrino mass matrix $M$ are (in principle) computable. If
such more  unconventional scenarios should appear as predictions, a more careful analysis of constraints from
oscillation experiments, electroweak decays and cosmological observations would be
 required to test consistency of such a string model with respect to the neutrino sector data.

\section{Conclusions and outlook}

In this paper we have explored the structure of
neutrino masses originating from certain string theory instanton effects
recently pointed out in the literature \cite{iu,bcw,isu}.
They appear in string compactifications in which the SM group
is extended by a $U(1)_{B-L}$  getting  a Stuckelberg mass.
 Our analysis has concentrated in the simplest class
of such instantons with a  $Sp(2)$ Chan-Paton symmetry. These instantons lead to
a certain flavour-factorised form for both, the $\nu_R$ mass matrix and the Weinberg operator.
A hierarchy of neutrino masses naturally appears from the different values of
the actions for the different contributing instantons.
For the case that the Weinberg operator gives rise to the leading contribution
to neutrino masses, 
we have shown how one can reproduce the experimental patterns for neutrino masses
and mixings under not very restrictive conditions on the
instanton amplitudes $c_a^{(r)}$. For particular directions of these
flavour vectors  $c_a^{(r)}$ one may reproduce, for example, the
structure of tri-bimaximal mixing. This is true both for normal and
inverted hierarchy cases, although the latter seems more unlikely
within the present scheme.  In the opposite case in which the see-saw mechanism gives
rise to the leading contribution, the structure of neutrino masses depends
strongly on the form of the Dirac neutrino mass matrix. In a simplified situation with
a diagonal Dirac mass matrix one can obtain, e.g., tri-bimaximal mixing if the
flavour vector coefficients $d_a^{(r)}$ align along certain flavour directions.
The often assumed situation with a diagonal $\nu_R$ mass matrix and mixing originated
in the Dirac sector is also possible.

A number of extra possibilities should be explored. Other classes of string instantons \cite{isu} with
Chan-Paton symmetries $O(1)$ and $U(1)$ in general do not lead to a factorised flavour
dependence of both $\nu_R$ masses and Weinberg operator. It would be interesting to explore
the phenomenological possibilities for these other classes of instantons.
From the string model building point of view, it would be important to learn more about the
structure and flavour dependence of the flavour vectors $c_a^{(r)}$ and $d_a^{(r)}$
in particular string compactifications. To do that a search for models with an extra
$U(1)_{B-L}$ gauge boson which becomes massive through a Stuckelberg is required.
Getting a neutrino spectrum consistent with experimental constraints would  be
a new important test of string models.

One assumption we have made is that the contribution to the leptonic mixing matrix from
the mass matrix of the charged leptons is small. This condition is satisfied in many well
motivated phenomenological
models, where there is only small mixing in the mass matrices of quarks as well as charged leptons.
 We note however that in general, large mixing can as well stem from the charged lepton sector (see e.g.\ \cite{Antusch:2004re}) or
from a combination of both, neutrino and charged lepton contributions.
The conditions derived in this letter can be readily generalised to these scenarios as well.
For the case of small mixing from the charged lepton sector,
 the charged lepton contributions can be treated as corrections to the
neutrino mixing angles and CP phases (see e.g.~\cite{Antusch:2005kw}).
The general conditions for consistency with neutrino data do not change due these small corrections. In the case that the charged lepton mixing matrix is CKM-like, i.e., small and dominated by a 1-2 mixing, and for small 1-3 mixing in the neutrino mass matrix, the neutrino mixing sum rule  $\theta_{12} - \theta_{13}\cos(\delta)\approx \theta^\nu_{12}$ \cite{Antusch:2005kw,King:2005bj,Masina:2005hf} holds between the measurable PMNS parameters $\theta_{12},\theta_{13},\delta$ and the theoretical prediction for the 1-2 mixing angle $\theta_{12}^\nu$ from the diagonalisation of the neutrino mass matrix. Thus, the prediction for the neutrino mixing angle $\theta_{12}^\nu$, which is directly connected to the string instantons, can be tested by precisely measuring $\theta_{13},\theta_{12}$ and $\delta$ in future neutrino experiments \cite{Antusch:2007rk}.

Regarding leptogenesis, there is one conceptually interesting fact:
All of leptogenesis would have its origin
in instantons. $\nu_R$ masses would come from string instantons, and the
transformation of lepton into baryon asymmetry would be due to $SU(2)_L$ gauge instantons.
Leptogenesis would proceed via the out-of-equilibrium decay of the right-handed
(s)neutrinos, and in the general case both, the other right-handed neutrinos as well
as the Weinberg operator would contribute to the decay asymmetries proportional to
their contribution to the neutrino mass matrix \cite{Antusch:2007km}. 
In this respect is worth noting that the flavour vectors $c_a^{(r)},d_a^{(r)}$
are in general complex and so will be the generated neutrino mass matrices.
It would be  interesting to explore in more detail whether (semi-)realistic string
constructions consistent with the low energy neutrino data could
also give rise to successful baryogenesis via leptogenesis.

\vspace*{1cm}

{\bf \large Acknowledgments}

We thank  F. Marchesano, B. Schellekens and A. Uranga
  for useful discussions.
T.M. thanks the Instituto de F\'{\i}sica Te\'orica
IFT-UAM/CSIC  for hospitality while
this work was being carried out.
This work has been partially supported by the European Commission under
the RTN European Program  MRTN-CT-2004-503369,
the CICYT (Spain), and the Comunidad de Madrid under project HEPHACOS,
P-ESP-00346.

%\newpage

\end{document}